\documentstyle[prl,aps,multicol,amssymb,epsfig]{revtex}
\title{Spontaneous emission of an atom in front of a mirror}
\author{Almut Beige\footnote{a.beige@mpq.mpg.de}, 
Jiannis Pachos, and Herbert Walther}
\address{Max-Planck-Institut f\"ur Quantenoptik, D-85748 Garching, Germany}
\date{\today}

\begin{document}

\maketitle
\draft

\begin{abstract}
\begin{center}
\parbox{14cm}
{Motivated by a recent experiment [J. Eschner {\it et al.}, Nature {\bf 413}, 
495 (2001)], we now present a theoretical study on the fluorescence
of an atom in front of a mirror. On the assumption
that the presence of the distant mirror and a lens imposes boundary
conditions on the electric field in a plane close to the atom,
we derive the intensities of the emitted light as a function of
an effective atom-mirror distance. The results obtained are in good
agreement with the experimental findings.}
\end{center}
\end{abstract}

\vspace*{0.2cm}
\noindent
\pacs{PACS: 42.50.Lc, 03.65.Yz, 42.50.Ct}

\begin{multicols}{2}
\section{Introduction}

One of the fundamental subjects in quantum optics is describing
the fluorescence from single atom sources. Different scenarios have been
considered, the simplest one referring to an atom in free space\cite{atom}.
The fluorescence of an atom can be altered for example by the presence
of other atoms inducing dipole-dipole interactions\cite{dipole},
by the presence of a mirror\cite{fermi,mirror,eber,shift} or by the
single mode of the electromagnetic field inside a
cavity\cite{cavity,Lange}. To investigate experimentally these
phenomena ion trapping technology has been employed and good
agreement with theoretical predictions has been found.

Theoretical models have been developed starting from the Hamiltonian that
describes the atom, the free radiation field and their interaction.
To predict the time evolution of an ensemble of atoms, master equations can
be derived by tracing over all possible photon states. Alternatively, it
can be assumed that the environment performs continuous measurements on the
free radiation field. This leads to a quantum trajectory description\cite{traj}
which is especially appropriate for analysing experiments with single atoms.
Examples are experiments measuring the statistics of macroscopic light and
dark periods\cite{dehmelt,behe2} and the spectrum of the light from a three-level
atom with a metastable state\cite{tamm,plenio}. A quantum jump approach was also
applied to calculate the spatial interference pattern of the photons spontaneously
emitted by two atoms\cite{schoen} which was observed experimentally
by Eichmann {\em et al.}\cite{Eichmann}.

Recently, an experiment was conducted by Eschner {\it et
al.}\cite{eschner,eschner2} to measure the fluorescence of a single
three-level barium ion kept at a fixed distance from a
mirror. Qualitative explanations were given for most of the effects
observed. A recent theoretical study by Dorner and Zoller\cite{Dorner}
provides a detailed description of the experimental setup considering
a two-level atom and a one-dimensional model of the free radiation
field. Special attention is paid to a regime of large
atom-mirror distances where intrinsic memory effects cannot be
neglected anymore. In contrast to this we present here an alternative
study with a full three-dimensional treatment of the free radiation
field where delay time effects are considered
negligible. Nevertheless, same qualitative effects resulting from the
presence of the mirror as in\cite{Dorner} are
predicted and good quantitative agreement with the experimental
findings\cite{eschner,eschner2} is achieved. An earlier experiment
by Drexhage\cite{Drexhage} in 1974 observed the fluorescence from
molecules deposited on mirrors.

\noindent \begin{minipage}{3.38truein}
\begin{center}
\begin{figure}[h]
\centerline{
\put(30,-10){$1$}
\put(140,18){$2$}
\put(88,115){$3$}
\put(52,92){$\Delta_1$}
\put(120,97){$\Delta_2$}
\put(155,5){$\omega_m$}
\put(20,50){$\Omega_1$}
\put(150,60){$\Omega_2$}
\put(70,50){$\Gamma_1$}
\put(95,60){$\Gamma_2$}
\epsffile{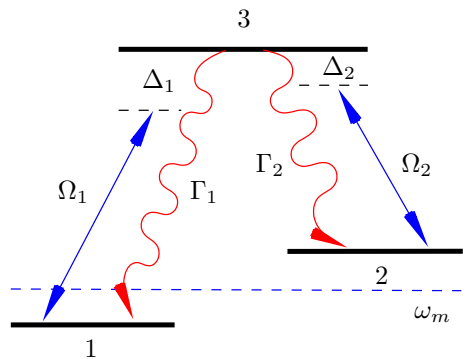}}
\vspace*{0.3cm}
\caption[contour]{\label{system} Atomic level scheme. Two lasers with Rabi
frequency $\Omega_1$ and $\Omega_2$ and detunings $\Delta_1$ and $\Delta_2$
drive the two transitions in the $\Lambda$ system. The free-space spontaneous
decay rates of the upper level are $\Gamma_1$ and $\Gamma_2$.}
\end{figure}
\end{center}
\end{minipage}
\vspace*{-0.5cm}

In experiment\cite{eschner}, the atom is driven by two detuned laser
fields and emits photons along 
two transitions that comprise a $\Lambda$ system (see Figure \ref{system}).
In the following, the Rabi frequency and the detuning of the laser field 
driving the 3-$j$ 
transition $(j=1,2)$ are denoted by $\Omega_j$ and $\Delta_j$, respectively, 
while $\Gamma_1$ and $\Gamma_2$ are the free-space spontaneous decay 
rates of the upper level. Detectors measure the intensities of 
the spontaneously emitted photons from the two transitions (see Figure \ref{mirror}). 
One detector is only sensitive to photons with
frequency $\omega_{31}$ and the measured intensity shows a strong
sinusoidal dependence on the atom-mirror distance $r$ with maximum
visibility of $72\,\%$. The other detector, measuring the photons with
frequency $\omega_{32}$, which are {\em not} affected by the mirror,
sees an intensity that depends only weakly on $r$ and has a
visibility of about $1\,\%$. The maxima of the two light intensities are
shifted with respect to each other.

\noindent \begin{minipage}{3.38truein}
\begin{center}
\begin{figure}[t]
\centerline{
\put(138,35){Atom}
\put(77,5){Mirror}
\put(-5,30){$\omega_{32}$ detector}
\put(180,30){$\omega_{31}$ detector}
\put(110,100){Lasers}
\put(110,68){$r$}
\epsffile{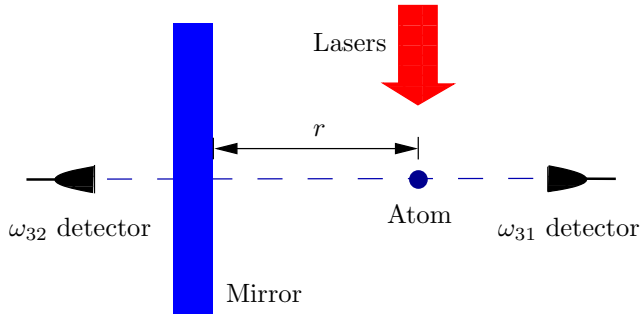}} 
\vspace*{0.3cm}
\caption[contour]{\label{mirror} Experimental setup of an atom placed 
at a fixed distance $r$ from a mirror and emitting photons with 
frequencies $\omega_{31}$ and $\omega_{32}$. The mirror is only sensitive to 
photons with frequency $\omega_{31}$ and two detectors measure 
light intensities.}
\end{figure}
\end{center}
\end{minipage}
\vspace*{-0.5cm}

Here it is assumed that the lens placed between the atom and the 
mirror in experiment\cite{eschner} projects the boundary conditions imposed by
the mirror on the free radiation field onto a plane close to the atom. 
As the atom was located near the focus point of the lens, the experiment 
can be described by the setup in Figure \ref{mirror} with an
effective atom-mirror distance of the order of the wavelength $\lambda_{31}$. 
The aim of this paper is to explain the experiment with a quantum mechanical 
approach. Qualitative and quantitative agreement with the experimental results 
is obtained.

The paper is organised as follows. Section II presents a quantum jump
description of an atom in front of a mirror. The spatial-dependent 
decay rates and level shifts of the atom are calculated and master equations 
are derived to find the steady 
state of the laser-driven atom. These are the ingredients necessary to 
calculate the intensities of the emitted photons. 
In Section III we apply our results to the experiment by Eschner 
{\it et al.}, while Section IV shows that many aspects of the experiment can also 
be predicted by means of a mirror-atom model resulting 
from a comparison of the setup with a classical analog. For example, the effect 
of the mirror, modifying the overall decay rate of the upper atomic level
and introducing an $r$-dependent level shift, can be understood as 
{\em subradiance} between the atom and its mirror image. 
Finally, an overview of the paper is presented in the conclusions.

\section{Quantum mechanical description}

In this section the atom in front of the mirror is described by the quantum jump
approach\cite{traj}. The latter consists of two main parts; on the one hand, 
it gives the time evolution of the atom when no photons are
emitted, and on the other, it gives the spatial distribution of the 
emitted photons depending on the particular state of the atom at the time 
of the emission. Let us see how this can be obtained from the Schr\"odinger 
equation. The corresponding Hamiltonian is of the form
\begin{equation} \label{h}
H = H_{\rm atom} + H_{\rm field} + H_{\rm laser} + H_{\rm int} ~.
\end{equation}
The first three terms are the interaction-free Hamiltonian of the atom, 
the free radiation field and the classical laser field, while the last term
\begin{eqnarray} \label{int}
H_{\text{int}} &=& e \, {\bf D} \cdot {\bf E}({\bf r}) 
\end{eqnarray}
describes the interaction of the atom with the
quantised electric field in the dipole approximation. Here ${\bf D}$
is the atom dipole operator
${\bf D}={\bf D}_{31}|3\rangle\langle 1|+{\bf D}_{32}|3\rangle\langle2|+
\text{h.c.}$ and ${\bf E}({\bf r})$ is the observable of the free
radiation field at the position ${\bf r}$ of the atom modified by
the presence of the mirror. Choosing the coordinate system such that the mirror 
surface corresponds to the $x=0$ plane, leads to the classical constraint that,
at $x=0$, the component of the electric field parallel to the mirror surface 
has to vanish, i.e. 
\begin{equation} \label{bound}
{\bf E}_{\|} ({\bf x}:x=0)=0~,
\end{equation} 
for all frequencies that see the mirror. This classical constraint gives at
the quantum level a modification on the electric field observable 
restricting the expectation value of its parallel component to zero on
the surface of the mirror\cite{kurt}. Consider the case where due to
the mirror the radiation from the 3-1 transition gets reflected and
hence satisfies the constraint (\ref{bound}), while the mirror is
transparent for photons from the 3-2 transition. To take this
into account a cut-off frequency $\omega_m$ is introduced that lies
between the typical frequencies $\omega_{31}$ and $\omega_{32}$ of the
$\Lambda$ system. The mirror is assumed to be transparent for all
frequencies below $\omega_m$, and perfectly reflective for frequencies
above it. As it is seen later the results derived in this section are
independent of the exact value of the chosen cut-off frequency. At an
arbitrary position ${\bf x} =(x,y,z)$ in the right
half space of the mirror (see Figure \ref{mirror}) and with 
${\bf k}=k_{\|} \, \hat{\bf k}_{\|} + k_x \, \hat{\bf x}$ the electric field 
observable can then be written as\cite{mirror}
\begin{eqnarray}\label{E}
{\bf E}({\bf x}) &=& {\rm i} \hspace*{-0.2cm} \sum _{{\bf k}\lambda : \,
\omega_k<\omega_m } \hspace*{-0.2cm}
\left( {\hbar \omega_k \over 2 \varepsilon_0 V} \right)^{1/2}
\text{\boldmath{$\epsilon$}}_{{\bf k} \lambda} \, a_{{\bf k} \lambda} \, 
{\rm e}^{{\rm i}{\bf k} \cdot {\bf x}} \nonumber \\ 
&& + {\rm i} \hspace*{-0.2cm} \sum_{ {\bf k}: \,
\omega_k\geq \omega_m } \hspace*{-0.2cm} \left( {\hbar \omega_k
\over \varepsilon_0 V} \right)^{1/2}
\Big[ \, {\rm i} \, \big( \hat{\bf x} \times \hat{\bf k}_{\|} \big) 
\, \sin k_x x  \, a_{{\bf k} 1} \nonumber \\
&& + {1 \over k} \big( \, k_{\|} \hat{\bf x} \cos k_x x 
- {\rm i} \, k_x \hat{\bf k}_{\|} \sin k_x x \big) \, a_{{\bf k} 2} \Big] \,
{\rm e}^{{\rm i} {\bf k}_{\|} \cdot {\bf x}} \nonumber \\ && +\text{h.c.} 
\end{eqnarray}
where $a_{{\bf k} 1}$ and $a_{{\bf k} 2}$ are the annihilation operators for
photons with polarisation 
$\text{\boldmath{$\epsilon$}}_{{\bf k} 1} 
= \hat{\bf x} \times \hat{\bf k}_{\|}$ and 
$\text{\boldmath{$\epsilon$}}_{{\bf k} 2} 
= (k_{\|} \hat{\bf x} - k_x \hat{\bf k}_{\|})/k$, respectively, and wave vector 
${\bf k}$.

From (\ref{int}) and (\ref{E}) the effect of the mirror on the
atomic fluorescence can be calculated. Assume
that the initial state of the atom is known and equals $|\psi \rangle$ 
while the free radiation field is in the vacuum state $|0_{\rm ph} \rangle$.
This is an allowed physical state that develops according to the Hamiltonian 
(\ref{h}) for a certain time $\Delta t$. If level 3 is populated, 
this time evolution leads to population of all possible one-photon 
states\cite{traj}. Consider now a detector placed in a certain direction
$\hat{\bf k}$ away from the atom that measures single photons resulting 
from the 3-$j$ transition\cite{schoen}. To determine the state of the system in case 
of a click at this detector one has to apply either the projector
\begin{equation} \label{p1}
I\!\!P_{\hat{\bf k}}^{(1)} = \hspace*{-0.2cm}
\sum_{ k \lambda : \, \omega_k \geq \omega_m }
\hspace*{-0.2cm} |1_{k \hat{\bf k} \lambda} \rangle
\langle 1_{k \hat{\bf k} \lambda} | \,\, ,
\end{equation}
if $j=1$, or the projector
\begin{equation} \label{p2}
I\!\!P_{\hat{\bf k}}^{(2)} = \hspace*{-0.2cm}
\sum_{ k \lambda : \, \omega_k < \omega_m }
\hspace*{-0.2cm} |1_{k \hat{\bf k} \lambda} \rangle
\langle 1_{k \hat{\bf k} \lambda} | \,\, ,
\end{equation}
if $j=2$. When a click is registered at a detector, the photon is  
absorbed and the free radiation field changes to its ground state 
$|0_{\rm ph}\rangle$. 

The probability density for a click can be obtained from the 
norm of the unnormalised state of the system after an emission and equals
\begin{equation} \label{I}
I^{(j)}_{\hat{\bf k}}(\psi) = \lim _{\Delta t\rightarrow 0}
\, {1 \over \Delta t} \,
\big \| \, I\!\!P_{\hat{\bf k}}^{(j)} \, U(\Delta t,0) \, |0_{\text{ph}} \rangle
|\psi \rangle \, \big \|^2 ~.
\end{equation}
If the coupling constants of the atom to the free radiation field are introduced  
as
\begin{eqnarray}
g^{(1)}_{{\bf k} \lambda} &\equiv& - e \, 
(\omega_k / \varepsilon_0 \hbar V )^{1/2} \, 
{\bf D}_{31} \cdot \text{\boldmath{$\epsilon$}}_{{\bf k}\lambda} \nonumber \\
g^{(2)}_{{\bf k} \lambda} &\equiv& {\rm i} e \, 
(\omega_k / 2 \varepsilon_0 \hbar V )^{1/2} \, 
{\bf D}_{32} \cdot \text{\boldmath{$\epsilon$}}_{{\bf k}\lambda}
\end{eqnarray}
and the dipole moment ${\bf D}_{31}$ is taken for
convenience\cite{footnote} to be parallel to the mirror surface,
the interaction Hamiltonian can, with respect to the interaction-free Hamiltonian 
and within the rotating wave approximation, be written as
\begin{eqnarray} \label{inter}
H_{\text{int}}^{\rm (I)} &=& 
\hbar \hspace*{-0.2cm}
\sum_{{\bf k}\lambda: \, \omega_k < \omega_m } \hspace*{-0.2cm}
g^{(2)}_{{\bf k}\lambda} \, a_{{\bf k}\lambda}\, {\rm e}^{{\rm i} {\bf k} \cdot {\bf r}} 
\, {\rm e}^{{\rm i}(\omega_{32}-\omega_k )t} \, |3 \rangle \langle 2|
\nonumber \\
&& + 
\hbar \hspace*{-0.2cm}
\sum_{ {\bf k}\lambda: \, \omega_k\geq \omega_m } \hspace*{-0.2cm}
g^{(1)}_{{\bf k}\lambda} a_{{\bf k}\lambda}
\, {\rm e}^{{\rm i} {\bf k}_{\|} \cdot {\bf r}}
\, {\rm e}^{{\rm i}(\omega_{31}-\omega_k )t} \, |3 \rangle \langle 1| 
\, \sin k_x r 
\nonumber \\ && + \text{h.c.} 
\end{eqnarray}
Using first-order perturbation theory and the approximations usually
applied in quantum optics, (\ref{I}) leads to
\begin{eqnarray} \label{emm1a}
I^{(1)}_{\hat{\bf k}}(\psi) &=& {3 \Gamma_1 \over 4 \pi} \,
\big( 1- |\hat{{\bf D}}_{31} \cdot \hat{\bf k}|^2 \big) \, P_3(\psi) \,
\sin ^2 k_{31x} r \nonumber \\&& 
\end{eqnarray}
for the photons that are affected by the mirror and 
\begin{eqnarray} \label{emm1b}
I^{(2)}_{\hat{\bf k}}(\psi) &=& {3 \Gamma_2 \over 8 \pi} \,
\big( 1- |\hat{{\bf D}}_{32} \cdot \hat{\bf k}|^2 \big) \,   P_3(\psi) 
\end{eqnarray}
otherwise. Here $\Gamma_j$
is the spontaneous emission rate of the atom in free space through the 3-$j$ 
channel, while $P_3(\psi) = |\langle 3|\psi \rangle|^2$ denotes the initial 
population in the excited state. This shows that the emission intensity of the 
3-1 transition strongly depends on the atom-mirror distance through its 
proportionality to the factor $\sin ^2 k_{31x} r$, while 
$I^{(2)}_{\hat{\bf k}}(\psi)$ is not a function of $r$.

It has hitherto been assumed that the atomic state $|\psi\rangle$ is always the same
by the time of an emission. This is not the case for the experimental 
setup in Figure \ref{mirror}, in which the atom is continuously driven by a laser field. 
To apply our results to this situation, the atom has to be described by the 
steady-state matrix $\rho^{\rm ss}$ and $P_3(\psi)$ has to be replaced by
$P_3(\rho^{\rm ss})= \langle3| \rho^{\rm ss} |3\rangle$. To calculate 
the stationary state master equations are employed. They are in general of the form
\begin{equation} \label{dot}
\dot \rho = - {{\rm i} \over \hbar} \, \big[ \, H_{\text{cond}} \, \rho 
- \rho \, H_{\text{cond}}^\dagger \, \big] + {\cal{R}}(\rho) ~.
\end{equation}
Here $H_{\rm cond}$ is the non-Hermitian Hamiltonian that describes the time evolution 
of the atom under the condition of no photon emission, while ${\cal{R}}(\rho)$ gives 
its unnormalised state after an emission. For the atom in front of a mirror,
${\cal{R}}(\rho)$ is given by
\begin{eqnarray} \label{R}
{\cal{R}}(\rho) &=& \sum_{j=1,2} \bar \Gamma_j \, 
|j \rangle \langle 3| \, \rho \, |3\rangle \langle j| ~,
\end{eqnarray}
where $\bar{\Gamma}_1$ and $\bar{\Gamma}_2$ are the modified overall decay rates. 
They are obtained by integrating $I^{(j)}_{\hat{\bf k}}(\psi)$ over all 
directions, which gives by definition $\bar \Gamma_j \, P_3(\psi)$. This leads
for the dipole moment ${\bf D}_{31}$ oriented parallel to the mirror, as 
in\cite{mirror}, to
\begin{eqnarray} \label{gamma2}
\bar \Gamma_1 &=& \Gamma_1 
\left[ \, 1 - {3 \over 2} \left( { \sin 2 k_{31} r
\over 2 k_{31} r } + {\cos 2 k_{31} r \over (2 k_{31} r)^2}- { \sin 2
k_{31} r \over (2 k_{31} r)^3 } \right) \right]  \nonumber \\ &&
\end{eqnarray}
and $\bar \Gamma_2 = \Gamma_2$.
As expected, the decay rate $\bar \Gamma_1$ is altered by the mirror
and (\ref{gamma2}) is in perfect agreement with the general case
presented in\cite{shift}.

To derive the conditional Hamiltonian $H_{\rm cond}$ we proceed as above and assume 
that the effect of the environment and the detectors is the same as the effect of 
continuous measurements on the free radiation field\cite{traj,schoen}. 
In case no photon is found after the time $\Delta t$, the projector 
onto the field vacuum $|0_{\rm ph} \rangle \langle 0_{\rm ph}|$ has to be applied 
to the state of the system. Thus we obtain
\begin{eqnarray}
|0_{\rm ph} \rangle \langle 0_{\rm ph}| U(\Delta t,0) |0_{\rm ph} \rangle |\psi \rangle
&\equiv& |0_{\rm ph} \rangle \, U_{\rm cond} (\Delta t,0)  |\psi \rangle ~. 
\nonumber \\&&
\end{eqnarray}
Using second-order perturbation theory and the same approximations as above, 
the no-photon time evolution is summarised within the Hamiltonian $H_{\rm cond}$, 
which is, in the Schr\"odinger picture, given by 
\begin{eqnarray} \label{hcond}
H_{\text{cond}}(t) &=& \sum_{j=1,2} 
{\textstyle{1 \over 2}} \hbar  \, \Omega_j \, {\rm e}^{{\rm i} 
(\omega_3-\omega_j-\Delta_j) t/\hbar} \, 
|j \rangle\langle 3| + {\rm h.c.} \nonumber \\ 
&& + \hbar \Delta \, |3\rangle\langle 3|
- {\textstyle{ {\rm i} \over2}} \hbar \, (\bar \Gamma_1 +\bar \Gamma_2) \, 
|3\rangle \langle 3| ~, 
\end{eqnarray}
where
\begin{eqnarray} \label{shift}
\Delta &=& {3\Gamma_1  \over 4} 
\left( { \cos 2 k_{31} r \over 2 k_{31} r } - {\sin 2 k_{31} r \over (2 k_{31} r)^2}
- { \cos 2 k_{31} r \over (2 k_{31} r)^3 } \right) 
\end{eqnarray}
is, in agreement with\cite{shift},  
the level shift of the excited atomic state $|3\rangle$ resulting from the 
modification of the free radiation field due to the presence of the mirror.

From (\ref{dot}), (\ref{R}) and (\ref{hcond}) and the condition 
$\dot{\rho}^{\rm ss}=0$ the expression for the steady-state population 
of the excited state, $P_3(\rho^{\text{ss}})$, is obtained:
\end{multicols}
\begin{minipage}{6.92truein}
\vspace*{-0.5cm}
\begin{figure}
\hspace*{-0.4cm} \epsfig{file=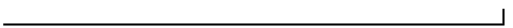,width=3.39truein} \\ 
\end{figure}
\begin{eqnarray} \label{p3}
P_3(\rho^{\text{ss}}) = && 4(\bar \Delta_1 -\bar \Delta_2)^2 (\bar
\Gamma_1 +\bar \Gamma_2) \Omega_1^2 \Omega_2^2 \,\, \nonumber \\ \nonumber \\ 
&& \times \Big\{ 
\big[(\Omega_1^2 + \Omega_2^2)^2 + 8 (\bar \Delta_1 - \bar \Delta_2)^2 \bar
\Gamma_1 \bar \Gamma_2 \big](\bar \Gamma_1 \Omega_2 ^2 + \bar \Gamma_2 \Omega_1 ^2 )
+ 4(\bar \Delta_1 - \bar \Delta_2)^2 \bar \Gamma_1 \bar \Gamma_2 
(\bar \Gamma_1 \Omega_1 ^2 + \bar \Gamma_2 \Omega_2 ^2 )
\nonumber\\ \nonumber \\
&& \hspace{0.5cm} +4 (\bar \Delta_1 -\bar \Delta_2)^2 
\left[ \bar \Gamma_1^3 \Omega_2^2 + 2 (\bar \Gamma_1 + \bar \Gamma_2) \Omega_1^2 \Omega_2^2
+ \bar \Gamma_2^3 \Omega_1^2 \right] 
- 8 (\bar \Delta_1 - \bar \Delta_2) (\bar \Delta_1 \bar \Gamma_1 \Omega_2^4 
- \bar \Delta_2 \bar \Gamma_2 \Omega_1^4 )  \nonumber \\ \nonumber \\ 
&& \hspace{0.5cm} 
+ 16 (\bar \Delta_1 - \bar \Delta_2)^2 \big( \bar \Delta_1^2 \bar \Gamma_1 \Omega_2^2 
+ \bar \Delta_2^2 \bar \Gamma_2 \Omega_1^2 \big) \Big\}^{-1} ~,
\end{eqnarray}
\vspace*{-0.2cm}
\begin{figure}
\hspace*{3.43truein} \epsfig{file=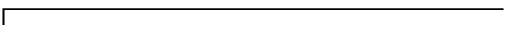,width=3.39truein} 
\end{figure}
\end{minipage}
\begin{multicols}{2}
\vspace*{-1.4cm} \noindent
where the notation $\bar \Delta_j \equiv \Delta_j-\Delta$ has been introduced.
This result shows that the stationary state of the atom is indeed affected by 
the presence of the mirror because of its dependence on the decay rate 
$\bar \Gamma_1$ and the level shift $\Delta$. Hence, both intensities 
$I^{(1)}_{\hat{\bf k}}(\rho^{\rm ss})$ 
and $I^{(2)}_{\hat{\bf k}}(\rho^{\rm ss})$ show spatial modulations 
originating from the boundary condition applied on the electric field 
observable ${\bf E}$. 

\section{Comparison with experimental results} 

In the experiment by Eschner {\it et al.}\cite{eschner}, a lens was employed 
to enhance the effect of the mirror in the neighborhood of the atom which was 
placed near the focus point $F$. The lens creates an image of the mirror near the 
atom, effectively changing the atom-mirror distance. With the same notation 
as in Figure \ref{refl} and classical optics considerations, it is  
seen that the distance between the mirror image and 
$F$ is $x=f^2/R$. Considering the distances used in the experiment, where
$f=12.5 \,{\rm mm}$ and $R=25\, {\rm cm}$, we obtain $x=625\, {\rm \mu
m}$. Since the atom is located close to $F$ it is also located very near
the mirror image. Alternatively, one might consider the geometrically 
equivalent model where the atom is projected by the lens into the 
neighborhood of the mirror with an effective distance $r$ from it
which can be made to be of the order of the wavelength
$\lambda_{31}=493\,{\rm nm}$\cite{rem}.

\noindent \begin{minipage}{3.38truein} 
\begin{center} 
\begin{figure}[h] 
\centerline{ 
\put(160,54){$F$}
\put(172,83){Mirror}
\put(173,73){image}
\put(113,82){Lens}
\put(75,5){$R$}
\put(139,5){$f$}
\put(175,5){$x$}
\epsffile{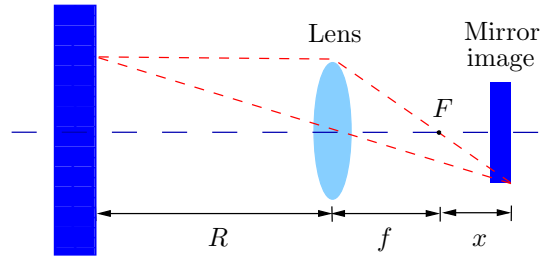}} 
\vspace*{0.3cm} 
\caption[contour]{\label{refl} The mirror and its image due to the 
presence of the lens. Here $R$ is the distance of the mirror from the lens, 
$f$ the distance from the lens to the focus point $F$, while $x$ is the distance 
of $F$ from the mirror image.} 
\end{figure} 
\end{center} 
\end{minipage} 
\vspace*{-0.5cm}

Boundary condition (\ref{bound}) also applies to the mirror image. 
In particular, the configuration of the electromagnetic field in the 
neighborhood close to the mirror surface is mapped on the
neighborhood around the atom. For suitable positions of the atom near
the mirror image and on the mirror-lens axis, the electromagnetic
field observable in its surrounding is faithfully given by (\ref{E}). 
Note that this consideration is also effective even when the solid angle 
with which the atom sees the lens is only $4\, \%$ as in experimental 
setup\cite{eschner}. Hence, the theoretical model considered in Section II 
should give the measured intensities assuming that the dipole moment of the atom 
was oriented parallel to the mirror surface.

Figure \ref{curves} shows the intensities $I^{(1)}_{\hat{\bf k}}(\rho^{\rm ss})$ 
and $I^{(2)}_{\hat{\bf k}}(\rho^{\rm ss})$ as a function of $r$ where the 
relevant parameters have been taken from\cite{eschner}. 
As expected, the photons which see the mirror show a very strong sinusoidal
$r$-dependence. If the effective atom-mirror distance $r$ is of the order 
of the wavelength $\lambda_{31}$, then the intensity measured by a 
detector behind the mirror also shows an $r$-dependence. Nevertheless, this 
dependence is much weaker and vanishes for large $r$. The relative order of 
magnitude of the intensities presented in Figure \ref{curves}, assuming 
$r \sim 5 \, \lambda_{31}$, is in agreement with the experimental findings 
(see Figure 3 in\cite{eschner}). 

\noindent \begin{minipage}{3.38truein} 
\begin{center} 
\begin{figure}[h] 
\centerline{ 
\put(0,220){(a)}
\put(0,90){(b)}
\put(120,-7){$k_{31}r$}
\epsffile{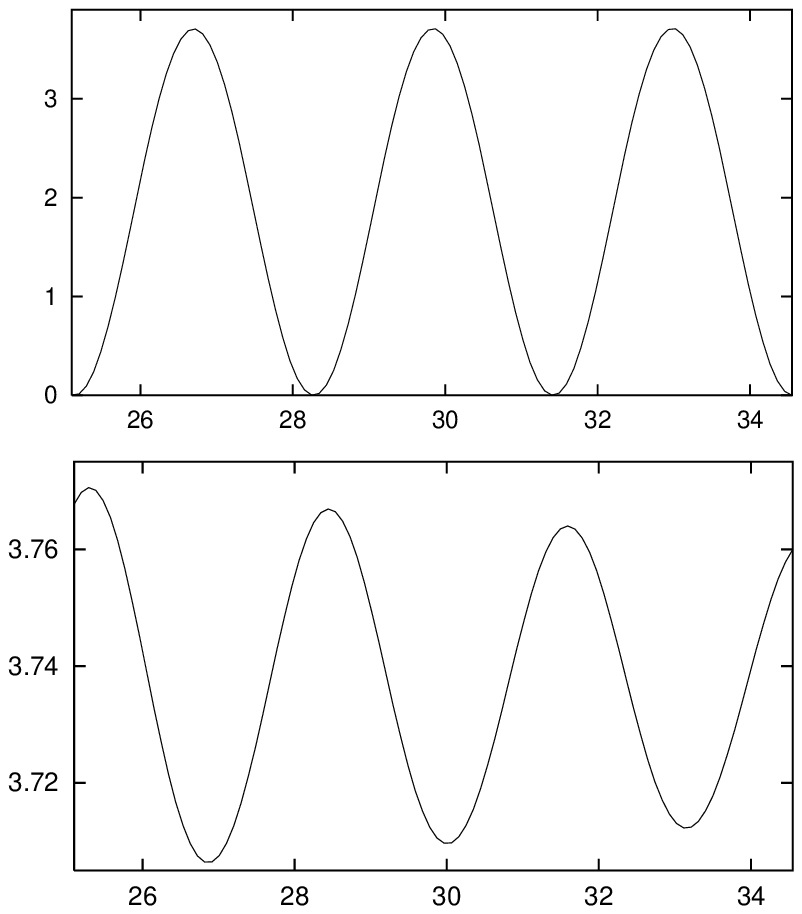}} 
\vspace{0.2cm}
\caption[contour]{\label{curves} The intensities, $I^{(1)}_{\hat{\bf
k}}(\rho^{\rm ss})$ (a) and $I^{(2)}_{\hat{\bf k}}(\rho^{\rm ss})$ (b) as a 
function of the effective atom-mirror distance for $\Delta_1=2 \,{\rm MHz}$, 
$\Delta_2=0$, $\Omega_1=10 \,{\rm MHz}$, $\Omega_2=5 \,{\rm MHz}$,
$\Gamma_1=15.1\,$MHz and $\Gamma_2=5.4\,$MHz. The vertical axis is given in 
units of $10^{-2}\,{\rm MHz}$ for both plots.}
\end{figure} 
\end{center} 
\end{minipage} 
\vspace*{-0.5cm} 

In addition, Figure \ref{curves} shows that the two intensities can be
anticorrelated, having a phase difference close to $\pi$. This effect
is due to their difference in nature and can vary for different values
of the $\Omega_i$ and $\Delta_i$ parameters. The origin of the pattern
in Figure \ref{curves}(a) is the $\sin^2 k_x r$ 
factor in $I^{(1)}_{\hat{\bf k}}(\rho^{\rm ss})$, while the $r$-dependence 
of the population $P_3(\rho^{\text{ss}})$ is in this case insignificant. 
In contrast, the pattern shown in Figure \ref{curves}(b) is only due to 
the $P_3(\rho^{\text{ss}})$-dependence of $I^{(2)}_{\hat{\bf k}}(\rho^{\rm ss})$. 
Its spatial configuration is dictated by $\bar \Gamma_1$, which includes the 
dominant term $\sin 2 k_{31} r$, and by $\Delta$, which includes the dominant term 
$\cos 2 k_{31} r$. Hence, the first plot is a
consequence of the modification of the electromagnetic field
observable in the neighborhood of the atom, while the second plot 
is a consequence of the modification of the spontaneous emission
rate $\bar \Gamma_1$ and the level shift $\Delta$ of the excited
atomic level. From this we can deduce that the anticorrelation of the
intensities takes place {\it only} for certain values of the
Rabi frequencies and detunings.

\noindent \begin{minipage}{3.38truein} 
\begin{center} 
\vspace{-5.8cm} 
\begin{figure}[ht] 
\centerline{
\put(90,300){$10^4 P_3$}
\put(250,195){$k_{31} r$}
\put(150,195){$\Omega_1$}
\epsffile{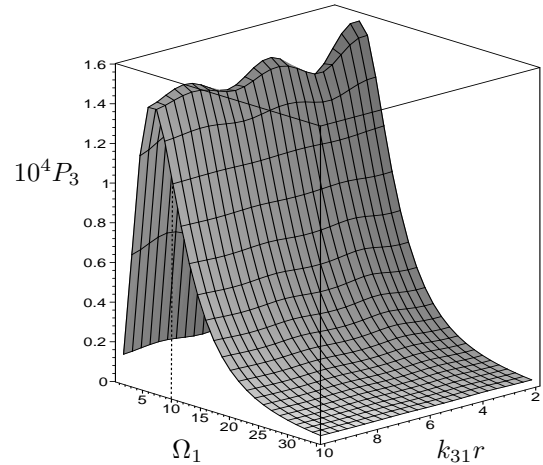}}
\vspace{-6cm} 
\caption[contour]{\label{3do1} 
Steady state population of the excited state, $P_3(\rho^{\text{ss}})$,
as a function of the Rabi frequency $\Omega_1$ (in MHz) and the
atom-mirror distance $r$ for $\Delta_1=0$, $\Delta_2=0.1\,$MHz and
$\Omega_2=10\,$MHz. For $\Omega_1=\Omega_2$ the spatial modulations
disappear (see dotted line). The $\pi$ phase change is apparent above
and below this value.}
\end{figure} 
\end{center} 
\end{minipage}

It is instructive to have a closer look at
the modulation of $P_3(\rho^{\text{ss}})$ as a function
of the Rabi frequencies and laser detunings. 
For example, for detunings $\Delta_1$ and $\Delta_2$ much smaller than
the Rabi frequencies of the driving laser fields relation (\ref{p3})
simplifies to
\begin{eqnarray} 
P_3(\rho^{\text{ss}}) &\approx&  
4(\Delta_1 - \Delta_2)^2 \,{\Omega_1^2 \Omega_2^2 \over (\Omega_1^2 +
\Omega_2^2)^2} \,{\bar \Gamma_1 +\bar \Gamma_2 \over  
\bar \Gamma_1 \Omega_2 ^2 + \bar \Gamma_2 \Omega_1 ^2} ~ .
\label{simple}
\end{eqnarray}
The second factor gives the main modulation of the intensity with
respect to the Rabi frequencies, while the third factor gives the
distance-dependent oscillations observed in Figure \ref{curves}(b).
The latter modulation
may change phase by $\pi$ if the ratio $\Omega_1/ \Omega_2$ changes
from smaller than one to larger than one, as can be predicted by
(\ref{simple}) and Figure \ref{3do1}. In particular, if
$\Omega_1 = \Omega_2$ then the modulations with $r$ vanish. On the
other hand, the maximum amplitude of the fringes appears for laser 
intensities for which also $I^{(2)}_{\hat{\bf k}}(\rho^{\rm ss})$ 
becomes maximal. For $\Delta_1=\Delta_2$, the
population $P_3(\rho^{\text{ss}})$ vanishes as a dark state is
generated between the levels $1$ and $2$. Trapping of the population
to a single ground state also occurs when one of the Rabi frequencies
becomes much larger than the other. 

Taking into account that experimentally a maximum visibility of $72 \, \%$ has 
been found for $I_{\hat{\bf k}}^{(1)}(\rho^{\rm ss})$, one can predict the 
reduction of the visibility $V$ for increasing Rabi frequencies from Figure \ref{3do1}.
In Eschner {\em et al.} it was argued that the reduction of the maximum
visibility from unity is mainly due to thermal motion of the ion, non-optimal 
cooling conditions, fluctuations of the atom-mirror distance and imperfect 
mapping of the mirror neighborhood to the neighborhood of the atom by the lens. 
In addition, it was observed that the visibility was greater than $ 50 \, \% $ 
for Rabi frequencies $\Omega_1$ below saturation, while it reduced to below 
$10\,\%$ when the Rabi frequency increased to 3-fold saturation. Indeed, 
from a figure similar to Figure \ref{3do1}, but for $\Omega_2 \sim 1$MHz, 
we see that at 3-fold saturation the population of level 3 and hence the amplitude 
of the oscillations of the intensity $I_{\hat{\bf k}}^{(1)}(\rho^{\rm ss})$ 
reduce by about $30$ times from their value at the saturation point. This can 
explain the $\Omega_1$-dependent reduction of the visibility observed 
experimentally.

\noindent \begin{minipage}{3.38truein} 
\begin{center} 
\vspace{-5.8cm} 
\begin{figure}[ht] 
\centerline{
\put(85,300){$10^2 P_3$}
\put(250,195){$k_{31} r$}
\put(150,195){$\Delta_1$}
\epsffile{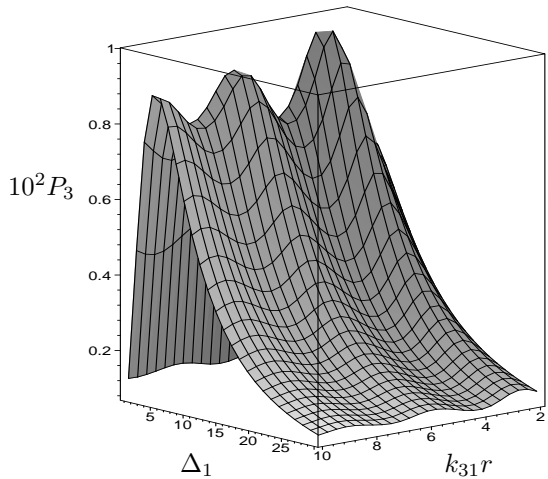}} 
\vspace{-6cm} 
\caption[contour]{\label{3dd1} 
Steady state population of the excited state, $P_3(\rho^{\text{ss}})$,
as a function of the detuning $\Delta_1$ (in MHz) and the atom-mirror
distance $r$ for $\Delta_2=0$, $\Omega_1=1\,$MHz and
$\Omega_2=10\,$MHz.}
\end{figure} 
\end{center} 
\end{minipage}

In Figures \ref{3dd1} we see the continuous phase change of the maxima of the 
population $P_3(\rho^{\rm ss})$ when the detuning $\Delta_1$ is varying. In 
particular, if we take the detunings much larger than the Rabi frequencies and 
decay rates, we obtain 
\begin{eqnarray}
P_3(\rho^{\text{ss}}) &\approx&  
{1 \over 4} \, \Omega_1^2 \Omega_2^2 \, {\bar \Gamma_1 +\bar \Gamma_2 \over  
\bar \Delta_1^2 \bar \Gamma_1 \Omega_2 ^2 + \bar \Delta_2^2 \bar \Gamma_2 \Omega_1 ^2} ~,
\end{eqnarray}
that gives an indication of the main terms which contribute to the continuous phase 
change of the $r$-dependent oscillations when one of the detunings is varying. 
Summarising this, by scanning different Rabi frequencies and detunings it 
should be possible to 
observe the variations in the amplitude of the population of level 3 as well as 
the discrete $\pi$ change or continuous change in the phase of the spatial 
modulations. In this way one could further verify our description of experimental 
setup\cite{eschner}.

\section{The mirror-atom model}

Describing the setup in Figure \ref{mirror} in a classical manner, one assumes 
that the atom is a point-like source with dipole characteristics.  
As it is classically possible to replace the mirror by a mirror-source 
at the distance $2r$, it could be assumed that the radiation properties of the 
atom can be predicted by replacing the mirror in the quantum setup by
a mirror-atom. 
Indeed, both descriptions lead to the same dependence of the light intensity on 
the source-mirror distance as found for $I^{(1)}_{\hat{\bf k}}(\psi)$ in 
(\ref{emm1a}). The mirror-atom model\cite{mirror} can 
even be used to predict further aspects of experiment\cite{eschner}. 
If the atom is initially prepared in the excited state $|3\rangle$ one has 
$P_3(\psi)=1$ and (\ref{gamma2}) gives the probability density for a photon 
emission. The quantum theory of dipole-interacting 
atoms is well known and a comparison with\cite{dipole} reveals that 
(\ref{gamma2}) coincides exactly with the decay rate of two dipole-interacting 
atoms prepared in the antisymmetric Dicke state of two two-level atoms at a 
distance $2r$. In addition, the level shift of $|3\rangle$ given in (\ref{shift}) 
equals the level shift of the antisymmetric state resulting from the dipole-dipole 
interaction.

Nevertheless, the mirror-atom model can no longer be used when 
the state of the atom by the time of the emission has no simple classical 
analog. It is not possible to take into account the driving of the two 
atomic transitions by a laser field. Dipole-interacting atoms have a richer 
structure of internal states and hence they cannot give completely 
equivalent results with the atom-mirror system. 

\section{Conclusions}

This paper presents a full quantum mechanical study of the fluorescence of 
an atom in front of a mirror, based on the assumption
that the mirror imposes boundary conditions on the electric field observable. 
In this way, the presence of the mirror affects the interaction of the atom 
with the free radiation field. This leads to a sinusoidal dependence 
of the intensities of the emitted light on the atom-mirror distance $r$. 
In addition, the overall decay rate of the atom becomes a function of $r$ 
and an $r$-dependent level shift is induced 
if $r$ is of comparable size to the wavelength of the emitted photons
-- an effect which can be interpreted in terms of {\em subradiance} due 
to dipole-dipole interaction between the atom and its mirror image.
 
In the actual experiment by Eschner {\em et al.}\cite{eschner}, 
the $25\,{\rm cm}$ distance between 
the atom and the mirror was much larger than the wavelength of the emitted 
photons and a lens was placed near the atom to enhance the effect of the
mirror. Motivated by this, a recent paper by Dorner and Zoller\cite{Dorner} 
took into account time-of-flight effects using a one-dimensional description 
of the free radiation field. Atom-mirror distances 
much larger than an optical wavelength were considered and delay differential 
equations were derived. Similar effects resulting from the presence of the 
mirror have been predicted, i.e. a sinusoidal dependence of the spontaneous decay 
rate and the level shift of the upper atomic level on the atom-mirror distance.
In contrast to the results presented here those modifications are persisting for 
large distances due to the one-dimensional character of their model
making it difficult to derive quantitative predictions comparable to the experimental 
findings.

In this paper it was assumed that the lens projects the mirror 
surface close to the atom so that the atom-mirror distance effectively becomes 
of similar size as the relevant wavelength. For the simplified setup, 
including only the atom and the mirror, a full three-dimensional description 
was given. Good qualitative and 
quantitative agreement was found with respect to different aspects of the 
experiment. Delay-time effects were neglected assuming that the relevant 
time scale for the projection of the mirror to the other side of the lens is 
in the experiment with about $1.7\,{\rm ns}$\cite{eschner} sufficiently smaller 
than the time scale on which the detector performs measurements on the free 
radiation field. This time scale has been denoted $\Delta t$ in Section II and 
is restricted from above only by the inverse decay rate of the relevant atomic 
transition\cite{traj} which equals $1/\Gamma_1 = 416\,{\rm ns}$.\\

{\em Acknowledgments.}
We would like to thank U. Dorner and B.-G. Englert for interesting and
helpful discussions. This work was supported by the European Union through 
IST-1999-13021-Project QUBITS.

\end{multicols}

\begin{references}
\bibitem{footnote} This choice of the orientation of the
dipole moment maximises the contrast of the observed fluorescence as a
function of $r$. The general case of the arbitrary dipole orientation
can be treated analogously\cite{mirror,shift,eber}, but we assume here that the
laser driving the 3-1 transition is chosen such that it excites only
an electric dipole moment oriented parallel to the mirror.
\bibitem{rem} Note that a small displacement of the atom corresponds
to an amplified displacement of the image of the atom near the mirror
by a few orders of magnitude. Because of this scaling the experiment
requires very efficient cooling of the atomic vibration.
\end{references}
\end{document}